\documentclass[12pt]{iopart}

\usepackage{iopams}
\usepackage{soul}
\usepackage{graphicx}
\usepackage{subfig}

\begin{document}

\title[]{Effects of Magnetic Helicity on 3D Equilibria and Self-Organized States in KTX Reversed Field Pinch}

\author{Ke Liu, Guodong Yu, Yuhua Huang, Wenzhe Mao$^*$, Yidong Xie, Xianyi Nie, Hong Li, Tao Lan, Jinlin Xie, Weixing Ding, Wandong Liu, Ge Zhuang and Caoxiang Zhu$^*$}
\address{School of Nuclear Science and Technology, University of Science and Technology of China, Hefei, 230026, People’s Republic of China}
\ead{maozhe@ustc.edu.cn and caoxiangzhu@ustc.edu.cn}
\vspace{10pt}
\begin{indented}
\item[]January 2023
\end{indented}

\begin{abstract}
The reversed field pinch (RFP) is a toroidal magnetic configuration in which plasmas can spontaneously transform into different self-organized states.
Among various states, the ``quasi-single-helical'' (QSH) state has a dominant component for the magnetic field and significantly improves confinement.
Many theoretical and experimental efforts have investigated the transitions among different states.
This paper employs the multi-region relaxed magnetohydrodynamic (MRxMHD) model to study the properties of QSH and other states.
The stepped-pressure equilibrium code (SPEC) is used to compute MHD equilibria for the Keda Torus eXperiment (KTX).
The toroidal volume of KTX is partitioned into two subvolumes by an internal transport barrier.
The geometry of this barrier is adjusted to achieve force balance across the interface, ensuring that the plasma in each subvolume is force-free and that magnetic helicity is conserved.
By varying the parameters, we generate distinct self-organized states in KTX. Our findings highlight the crucial role of magnetic helicity in shaping these states.
In states with low magnetic helicity in both subvolumes, the plasma exhibits axisymmetric behavior.
With increasing core helicity, the plasma gradually transforms from an axisymmetric state to a double-axis helical (DAx) state and finally to a single-helical-axis (SHAx) state. 
Elevated core magnetic helicity leads to a more pronounced dominant mode of the boundary magnetic field and a reduced core magnetic shear.
This is consistent with previous experimental and numerical results in other RFP devices.
We find a linear relationship between the plasma current and helicity in different self-organized states.
Our findings suggest that KTX may enter the QSH state when the toroidal current reaches 0.72 MA.
This study demonstrates that the stellarator equilibrium code SPEC unveils crucial RFP equilibrium properties, rendering it applicable to a broad range of RFP devices and other toroidal configurations.

\end{abstract}




\section{Introduction}

In recent decades, numerous magnetic confinement configurations have been proposed for nuclear fusion, with the tokamak and the stellarator emerging as the most widely recognized. 
The reverse field pinch (RFP), investigated as an alternative concept, features a toroidal magnetic field at the plasma edge that is opposite to the core.
It has the advantage of using ohmic heating to achieve fusion conditions \cite{martin2003overview, marrelli2021reversed}.
In contrast to stellarators, the RFP primarily generates its magnetic field through the plasma itself, without heavy reliance on external coils.
The poloidal field of RFP has a comparable order of magnitude to the toroidal field, which is different from that of a tokamak. 
The RFP plasma exhibits diverse states due to the presence of abundant self-organized effects.

In the axisymmetric equilibrium state, the safety factor $q$ of RFP is strongly sheared with $q(0)>0$ at the magnetic axis and $q(a)\leq0$ at the plasma edge. 
There are a large number of $m=1$ rational surfaces in the entire region. 
These rational surfaces are very unstable and it is easy for magnetic islands to emerge. 
These magnetic islands may overlap and turn the plasma core into chaos \cite{escande2000chaos, biewer2003electron}.
Destruction of nested magnetic surfaces reduces particle confinement capabilities. 
Experimentally, this state is characterized by the presence of many $m=1$ Fourier components of the magnetic field with comparable amplitude.

A self-organized helical equilibrium could become a new paradigm for ohmically heated fusion plasma \cite{escande2000quasi,lorenzini2009self}.
By increasing the plasma current, a spontaneous transition to a helical equilibrium occurs, with a high-temperature region in the core.
A $m=1$ and $n>3R/2a$ Fourier component of the magnetic field grows up and becomes dominant over the other secondary modes, where $a$ and $R$ are the minor and major radius of RFP configuration. This state is called the quasi-single-helical (QSH) state.

Two types of QSH states with distinct topologies have been observed \cite{lorenzini2008single, lorenzini2009improvement}. 
When the dominant component of the magnetic field is up to a few percent of the average magnetic field, there are two O-points in the magnetic field and two twisted magnetic axes. This state is known as the double-axis (DAx) state. 
As the amplitude of the dominant component increases, a new magnetic topology, which is characterized by a single helical magnetic axis has been experimentally observed. This new state is called the single-helical-axis (SHAx) state.

Taylor's relaxation theory can explain many plasma phenomena in RFP, including the reversed toroidal field \cite{taylor1974relaxation, taylor1986relaxation}.
According to Taylor's theory, the final state of relaxation is the state with minimum energy subject only to the single invariant
\begin{equation}
    \label{eq:helicity}
    \mathcal{H} = \int_{\mathcal{V}}\mathbf{A}\cdot\mathbf{B}\mathrm{d}\mathcal{V} \ ,
\end{equation}
where $\mathbf{A}$ is vector potential such that $\mathbf{B}=\nabla\times\mathbf{A}$ and $\mathcal{H}$ is the total magnetic helicity.
For a plasma enclosed by a perfectly conducting toroidal shell, the corresponding equilibrium of this state satisfies
\begin{equation}
    \label{eq:Beltrami field}
    \nabla\times\mathbf{B} = \mu\mathbf{B} \ ,
\end{equation}
where $\mu$ is a constant.
As a consequence, the plasma current is everywhere parallel to the magnetic field, and the pressure gradient $\nabla p$ vanishes.
Such a force-free field, meaning no pressure gradient force, can arise in both laboratory and space plasma.
If we ignore the toroidal curvature, the axisymmetric solution of \Eref{eq:Beltrami field} can be expressed using the Bessel function, 
\begin{equation}
   B_z=B_0J_0(\mu r)  \ ,
\end{equation}
\begin{equation}
   B_\theta=B_0J_1(\mu r)  \ ,
\end{equation}
where $r, \theta, z$ are cylinder coordinates. 
Reversal of the toroidal field will occur and $B_z(a)<0$ when $\mu a>2.404$.

While Taylor's theory has achieved significant success, experimental results show that the plasma in the RFP is not a strict global force-free equilibrium.
The ratio of parallel current to magnetic field $\lambda=\mu_0j_\parallel/B$ should be a strict constant in a force-free field.
However, the measurement of the current density in MST indicates that the $\lambda$ of the plasma core and edge are different \cite{brower2002measurement}.
The experiments of RFX-mod show that there is an electron transport barrier, and the gradients of various physical parameters are large near this barrier \cite{gobbin2011vanishing}.

Classic Taylor's theory cannot describe the 3D phenomena in RFP, so the stellarator codes have become an important tool for studying physics without axial symmetry in RFP. 
Gobbin et al. used the stellarator code VMEC for the first time to simulate the three-dimensional equilibrium in RFX-mod with the assumption of nested magnetic surfaces \cite{gobbin2011stellarator}.
There are some investigations of 3D characteristics and dynamo effect based on VMEC equilibrium in KTX \cite{liu2021QSH, Zu2022dynamo}. 
Furthermore, some studies show that the equilibrium of RFP without nested surfaces can also be reproduced \cite{qu2020flow, zhang2023}.

The multi-region relaxed magnetohydrodynamics (MRxMHD) is a new fluid model that combines the ideal MHD and Taylor relaxation \cite{hudson2012computation}. 
MRxMHD considers the helicity constraint of different subvolumes instead of the global constraint. 
Dennis et al. pointed out that the use of this model requires only a finite number of constraints to describe the self-organized plasma states in RFP \cite{dennis2013minimally}. 
However, this has not been fully investigated, and the impact of these constraints on the self-organized state of plasma has not been shown. 
A multiple domains scheme (MDS) is proposed to investigate the transport problem in the presence of magnetic islands, which also divides the plasma into multiple regions \cite{PhysRevLett.82.1458}. 
The MDS was applied to RFP plasma, which has been implemented in a new numerical tool \cite{Auriemma_2018}.

Here, we employ a barrier as the interface to divide the toroidal volume into inner and outer subvolumes.
Different self-organized states, including axisymmetric, DAx, and SHAx, can be reproduced numerically with the MRxMHD model. 
Helicity plays a critical role in this model and magnetic topology changes with helicity.

This paper is structured as follows.
We begin with an introduction to the variational principle of MRxMHD in \Sref{MRxMHD}. 
Details of the 3D equilibrium simulation of the Keda Torus eXperiment (KTX) are presented in \Sref{simulation}. 
There are more magnetic field properties of different self-organized states in \Sref{field}. 
The topological properties and the decomposition of helicity are shown in \Sref{decomposition}. 
We summarize in \Sref{summay}.

\section{Multi-region relaxed magnetohydrodynamics} \label{MRxMHD}

The energy functional of the ideal MHD is given by the integral
\begin{equation}
    \label{eq:energy}
    \mathcal{W} = \int_{\mathcal{V}}\left(\frac{p}{\gamma-1}+\frac{B^2}{2\mu_0}\right)\mathrm{d}\mathcal{V} \ ,
\end{equation}
where $\gamma$ is the adiabatic index and the plasma volume $\mathcal{V}$ is bounded by a toroidal surface.
The ideal MHD equilibrium equation $\mathbf{J}\times\mathbf{B}=\nabla p$ is obtained when \Eref{eq:energy} is extremized by using variational principle,  \cite{kruskal1958equilibrium, helander2014theory}
\begin{equation}
    \delta\mathcal{W} = 
    \int_{\mathcal{V}}(\nabla p - \mathbf{J}\times\mathbf{B})\cdot\xi\mathrm{d}\mathcal{V} \ ,
\end{equation}
where ideal displacement $\xi$ vanishes on the plasma boundary.

In Taylor's theory, the magnetic energy functional imposing constraint of the total magnetic helicity with a Lagrange multiplier $\alpha$ is expressed as \cite{taylor1974relaxation, imbert2019introduction}
\begin{equation}
    \label{eq:Taylor functional}
    \mathcal{F}_{\mathrm{Taylor}} 
    = \int_{\mathcal{V}}\left(\frac{B^2}{2\mu_0}\right)\mathrm{d}\mathcal{V}
    + \alpha\left(\int_{\mathcal{V}}\mathbf{A}\cdot\mathbf{B}\mathrm{d}\mathcal{V}-\mathcal{H}_0\right) \ .
\end{equation}
The first variation in $\mathcal{F}_{\mathrm{Taylor}}$ with respect to $\mathbf{A}$ is
\begin{equation}
    \delta\mathcal{F}_\mathrm{Taylor}  
    = \int_\mathcal{V}\delta\mathbf{A}\cdot\left(\frac{\nabla\times\mathbf{B}}{\mu_0}-2\alpha\mathbf{B}\right)\mathrm{d}\mathcal{V} \ .
\end{equation}
When $\delta\mathcal{F}_\mathrm{Taylor}=0$, the field satisfies \Eref{eq:Beltrami field} with $\mu=2\alpha/\mu_0$.

The first step towards constructing the MRxMHD energy functional in RFP is to separate the space into two subvolumes by an interface $\mathcal{I}$. 
In each subvolume, the magnetic helicity $\mathcal{H}_i$ is conserved, where $i=1,2$ represents the inner and outer subvolume respectively.
The MRxMHD energy principle aims to minimize the plasma energy subject to these constraints, 
\begin{equation}
    \mathcal{F} = \sum_i\mathcal{F}_i =  
    \sum_i\left(\mathcal{W}_i-\frac{\mu_i}{2}\left(\int_{\mathcal{V}_i}\mathbf{A}\cdot\mathbf{B}\mathrm{d}\mathcal{V}_i-\mathcal{H}_i\right)\right) \ ,
\end{equation}
where $\mathcal{W}_i$ is the energy given in \Eref{eq:energy} and $\mu_i$ is the Lagrange multiplier.

The first variation in the local constrained functional $\mathcal{F}_i$ due to the arbitrary variations $\delta\mathbf{B}=\nabla\times\delta\mathbf{A}$ in the field and the arbitrary variations $\xi$ in the interface $\mathcal{I}$ is given by \cite{hudson2012computation}
\begin{equation}
    \delta\mathcal{F}_i 
    = \int_{\mathcal{V}_i} (\nabla\times\mathbf{B}-\mu_i\mathbf{B})\cdot\delta\mathbf{A} \mathrm{d}\mathcal{V}_i
    - \int_{\partial\mathcal{V}_i}\left(p_i+\frac{B^2}{2}\right)\xi\cdot\mathrm{d}\mathbf{s} \ .
\end{equation}
When $\mathcal{F}$ is extremized, the magnetic field in each $\mathcal{V}_i$ follows the Beltrami equation,
\begin{equation}
    \label{eq:in subvolume}
    \nabla\times\mathbf{B} = \mu_i\mathbf{B} \ ,
\end{equation}
and the total pressure must be continuous across the interface $\mathcal{I}$, 
\begin{equation}
    \label{eq:across interface}
    \left[\left[p+\frac{B^2}{2}\right]\right]_\mathcal{I} = 0 \ ,
\end{equation}
where the bracket $[[*]]$ indicate the jump of the scalar across interface $\mathcal{I}$.

The stepped pressure equilibrium code (SPEC) serves as a numerical solver for the MRxMHD model \cite{hudson2012computation, hudson2020free, baillod2021computation}.
In SPEC, there could be any number of nested toroidal interfaces and the subvolume $\mathcal{V}_i$ enclosed by $\mathcal{I}_{i-1}$ and $\mathcal{I}_i$. 
The total pressure across $\mathcal{I}_i$ satisfies \Eref{eq:across interface} and the magnetic field in each $\mathcal{V}_i$ follows \Eref{eq:in subvolume}. 
Therefore, the pressure gradient $\nabla p$ is zero in each subvolume and the global pressure profile shows a stepped shape. 
SPEC solves the magnetic field of each subvolume $\mathcal{V}_i$ by iteratively adjusting the shape of the interface until the total force across the interface $\mathcal{I}_i$ is balanced.

\section{Self-organized equilibrium simulation of KTX} \label{simulation}

The KTX with a measurement system of helicity flux density located at the University of Science and Technology of China is a large RFP device \cite{liu2017overview, 10.1063/5.0073486}.
The KTX vacuum chamber has a circular cross-section with major radius $R=1.4\ \mathrm{m}$ and minor radius $a=0.4\ \mathrm{m}$.

\begin{figure}
    \centering
    \includegraphics[width=0.7\linewidth]{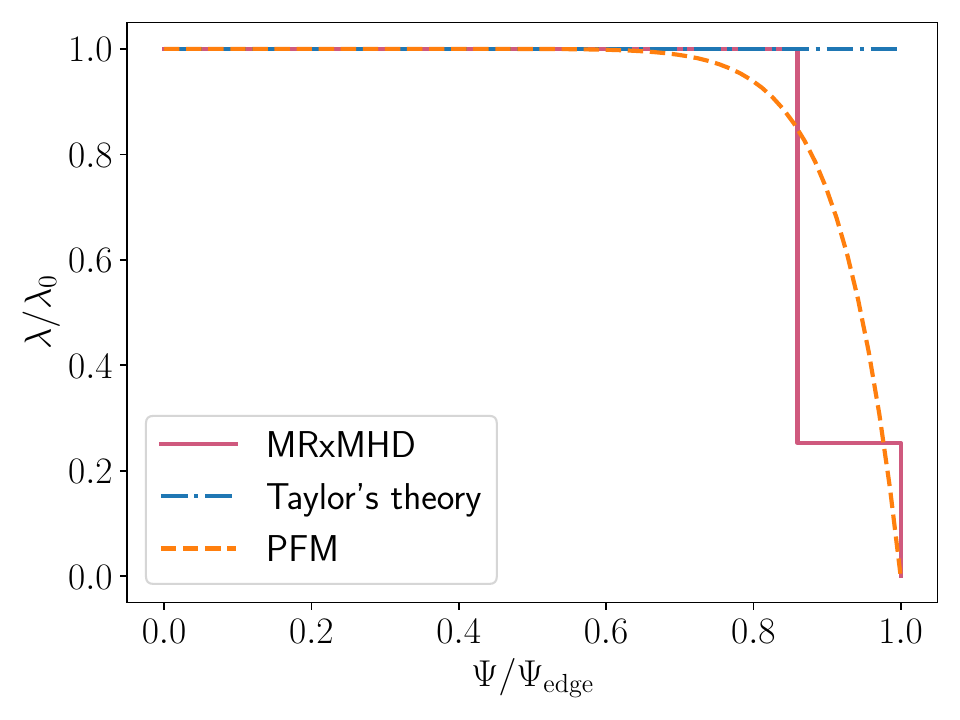}
    \caption{Radial profiles of $\lambda$ in different models.}
    \label{fig:lambda}
\end{figure}

We use SPEC to simulate the three-dimensional equilibrium in KTX. 
There are two subvolumes and an interface in the simulation. 
The state of the magnetic field in each subvolume is the state of minimum energy with invariant helicity, respectively. 
The magnetic field is obtained by solving \Eref{eq:in subvolume} in SPEC and choosing an appropriate interface shape to balance the total pressure across $\mathcal{I}$. 
According to Taylor's theory, the entire plasma should be in force-free equilibrium and $\lambda$ is a constant globally.
This is unattainable at the plasma edge. 
Although a polynomial function model (PFM) with $\lambda=\lambda_0(1-(r/a)^\alpha)$ is proposed \cite{Antoni_1986}, our simulation based on the MRxMHD uses two radially discontinuous $\lambda$ profiles to solve this problem.
\Fref{fig:lambda} illustrates the radial $\lambda$ profiles in different models.

\begin{figure}[htbp]
    \centering
    \subfloat[Axisymmetric State]{
        \label{fig:Axisymmetric State}
        \includegraphics[width=0.45\linewidth]{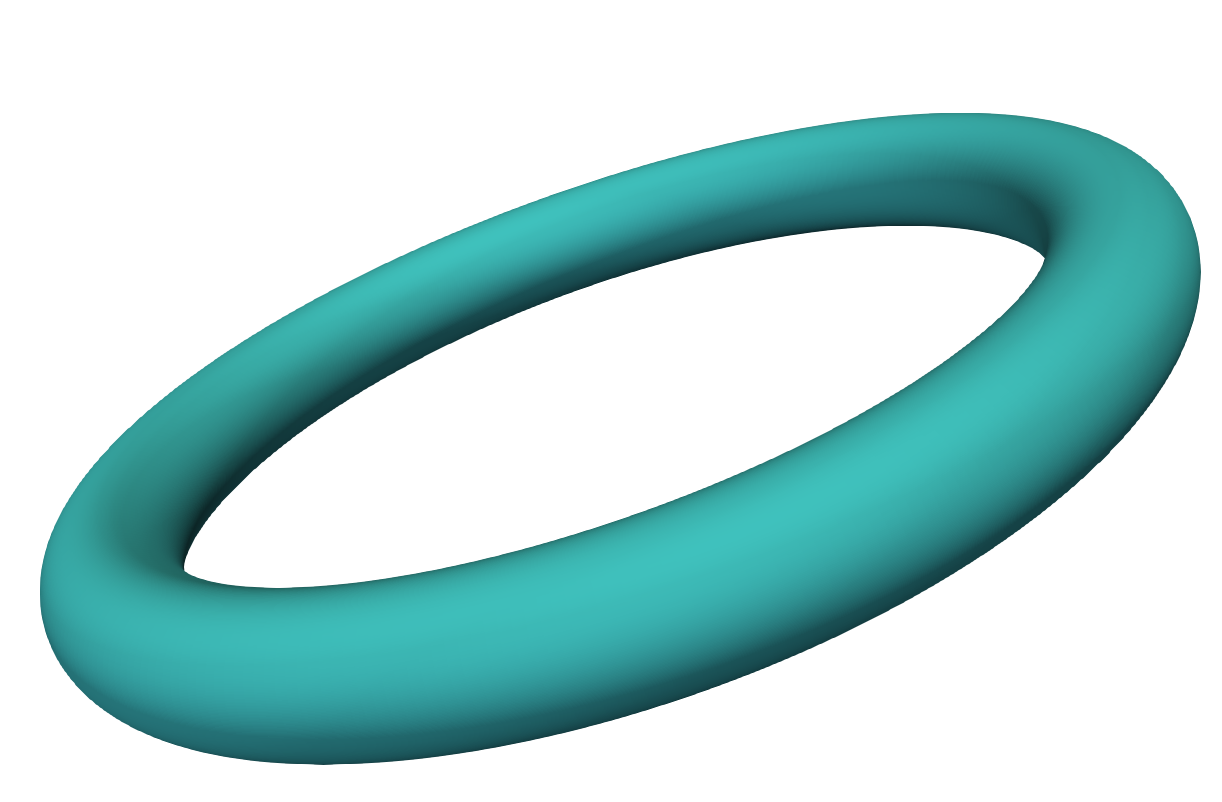}
        }\hfill
    \subfloat[DAx State]{
        \label{fig:DAx State}
        \includegraphics[width=0.45\linewidth]{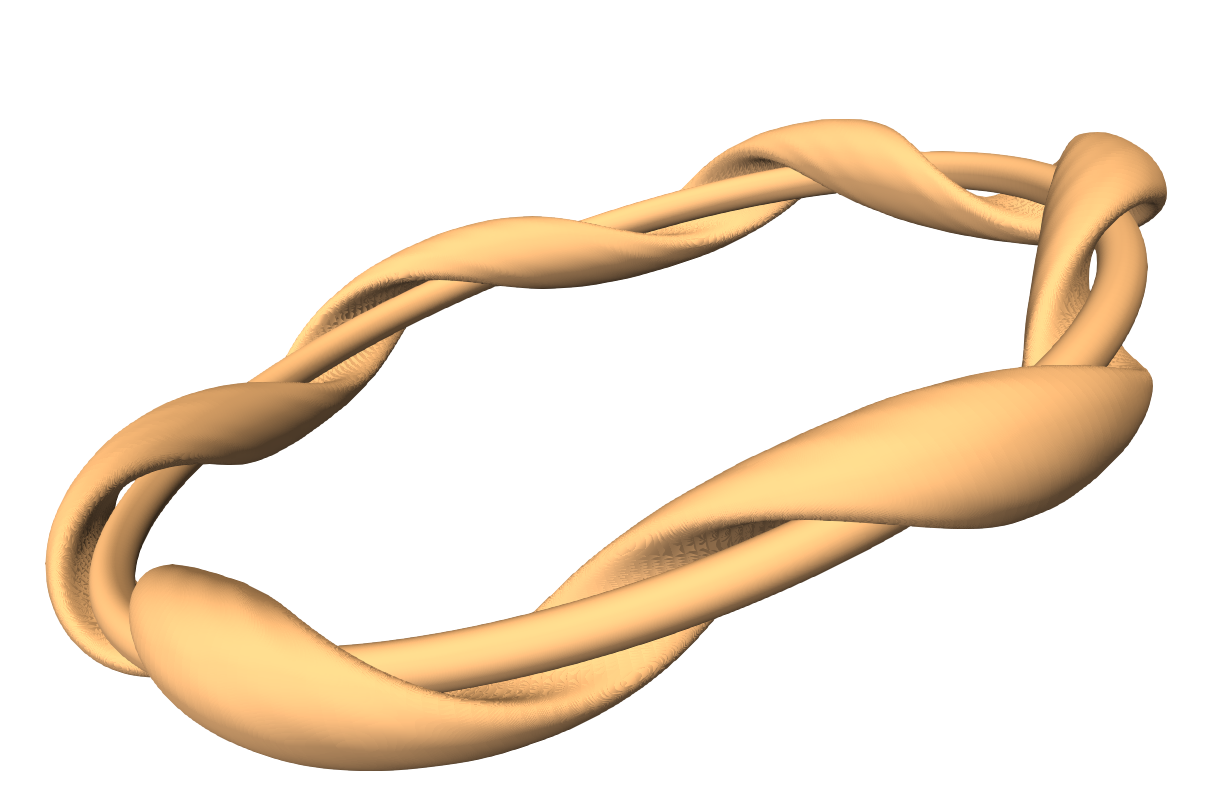}
        }\hfill
    \subfloat[SHAx State]{
        \label{fig:SHAx State}
        \includegraphics[width=0.45\linewidth]{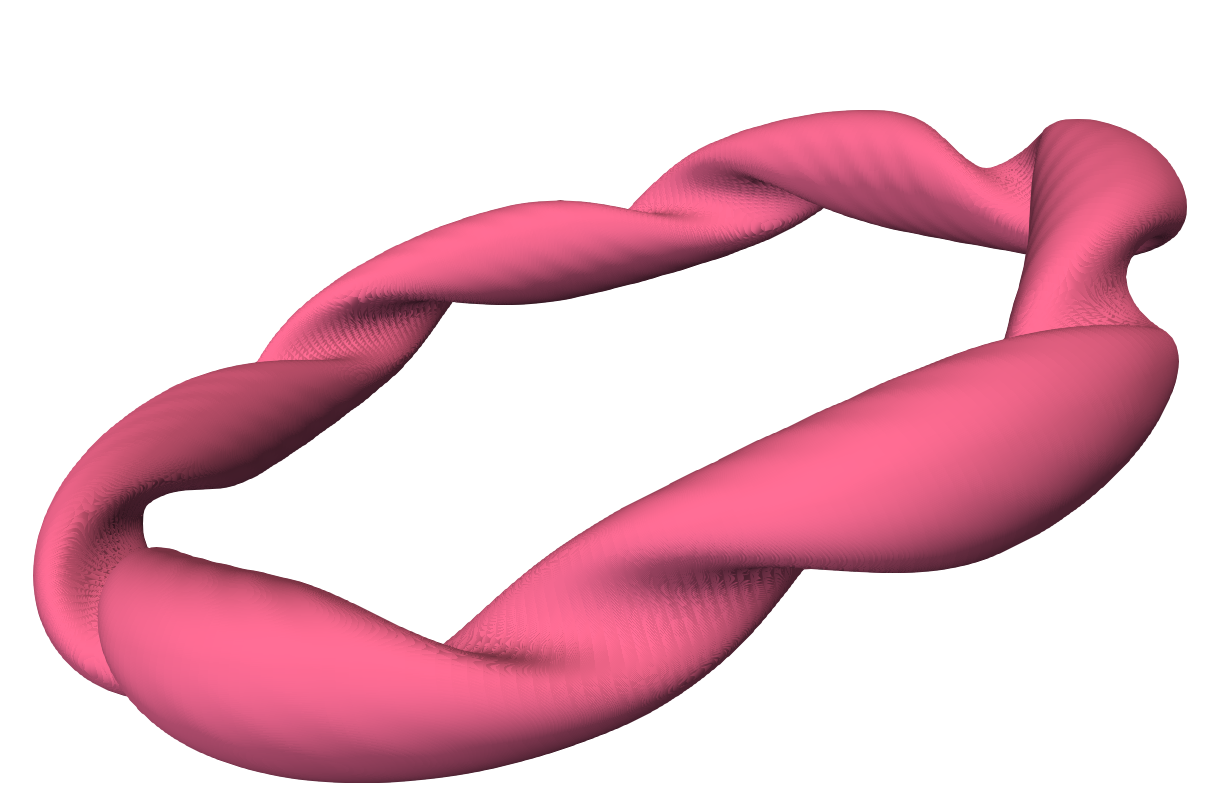}
        }
    \caption{Flux surfaces of different self-organized states. }
    \label{fig:flux surface}
\end{figure}

\begin{figure}
    \centering
    \subfloat[$\left<\mathcal{H}\right>=5.56\times10^{-3}$]{
        \label{fig:case 1}
        \includegraphics[width=0.45\linewidth]{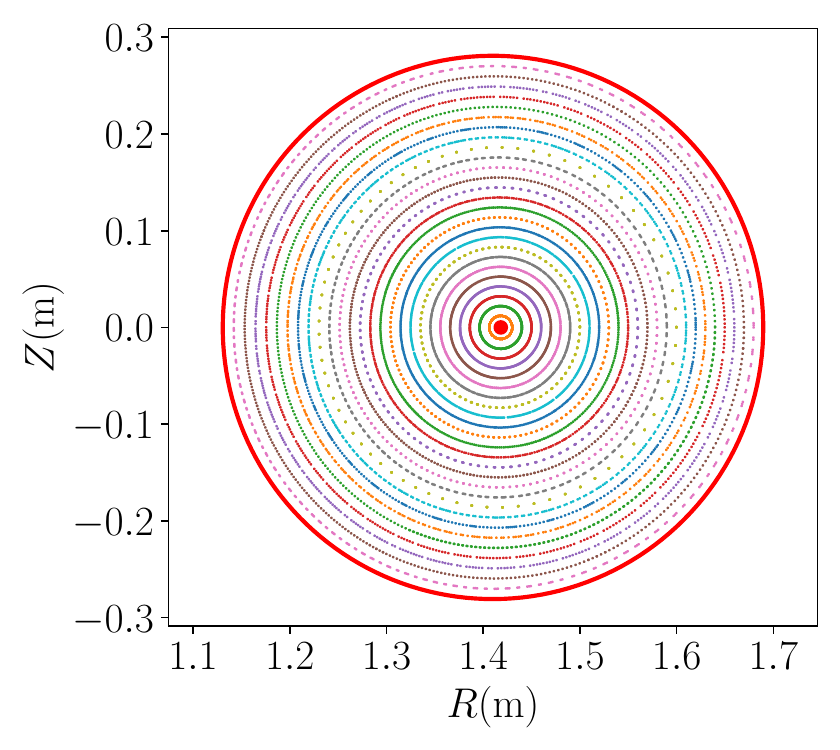}
        }\hfill
    \subfloat[$\left<\mathcal{H}\right>=6.64\times10^{-3}$]{
        \label{fig:case 2}
        \includegraphics[width=0.45\linewidth]{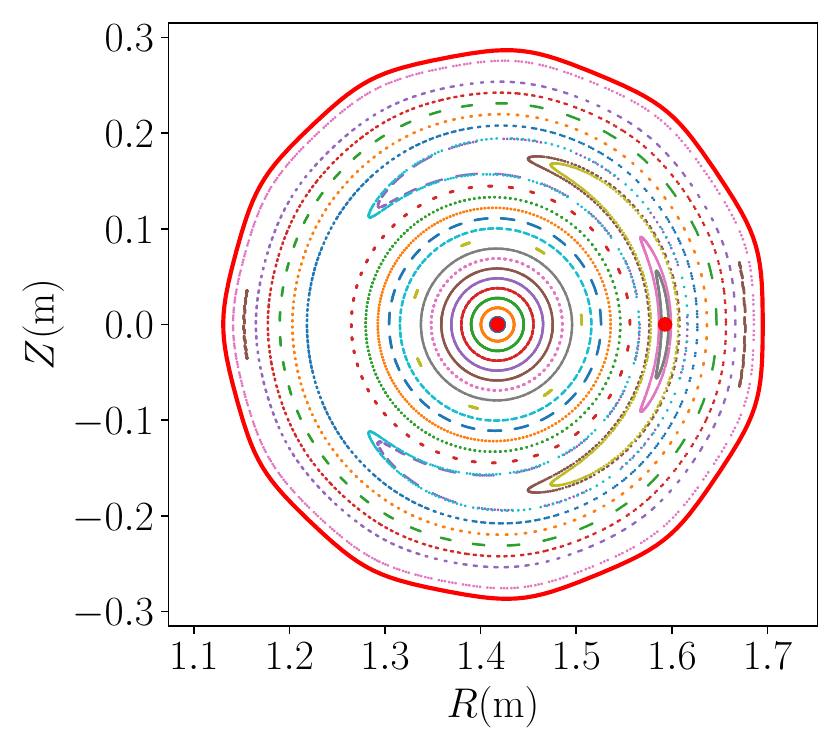}
        }\hfill
    \subfloat[$\left<\mathcal{H}\right>=7.92\times10^{-3}$]{
        \label{fig:case 3}
        \includegraphics[width=0.45\linewidth]{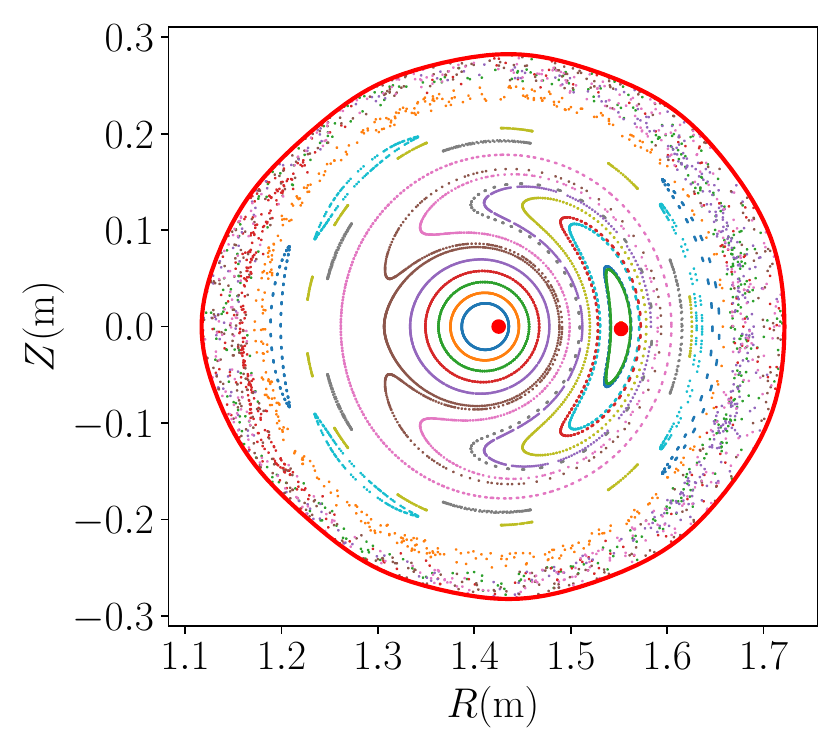}
        }\hfill
    \subfloat[$\left<\mathcal{H}\right>=8.41\times10^{-3}$]{
        \label{fig:case 4}
        \includegraphics[width=0.45\linewidth]{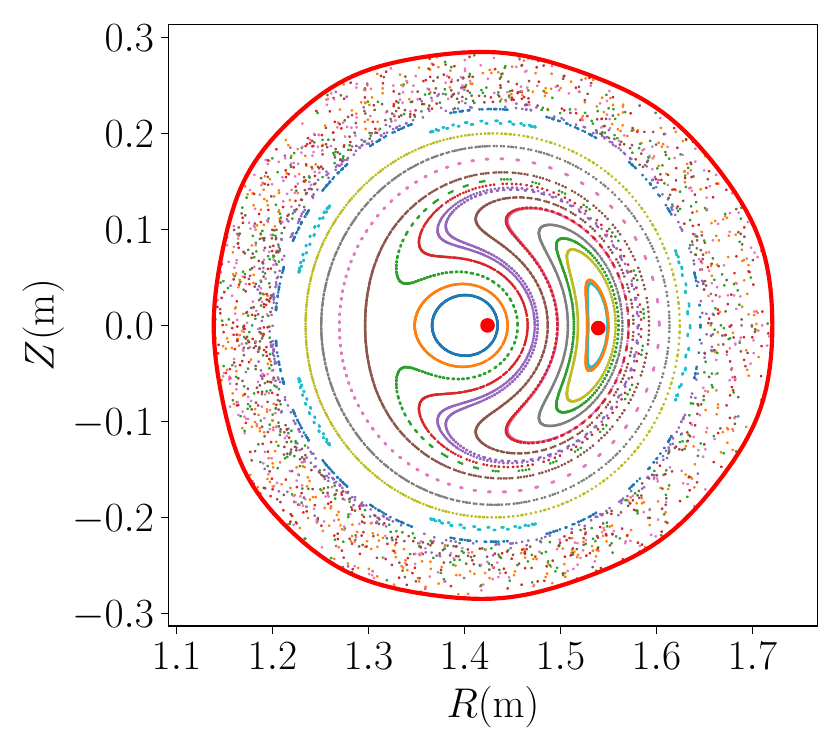}
        }\hfill
    \subfloat[$\left<\mathcal{H}\right>=8.72\times10^{-3}$]{
        \label{fig:case 5}
        \includegraphics[width=0.45\linewidth]{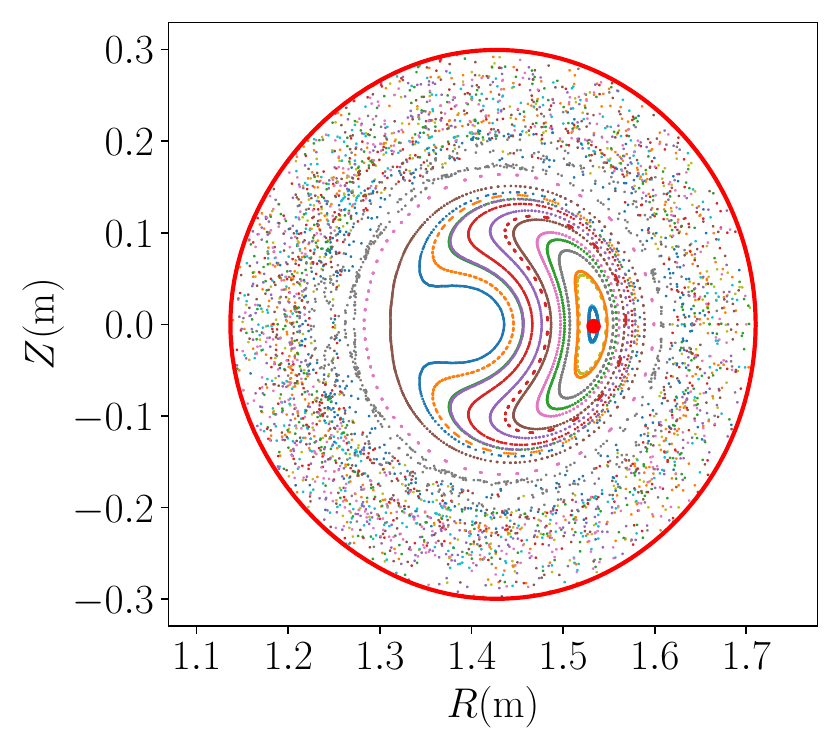}
        }\hfill
    \subfloat[$\left<\mathcal{H}\right>=9.03\times10^{-3}$]{
        \label{fig:case 6}
        \includegraphics[width=0.45\linewidth]{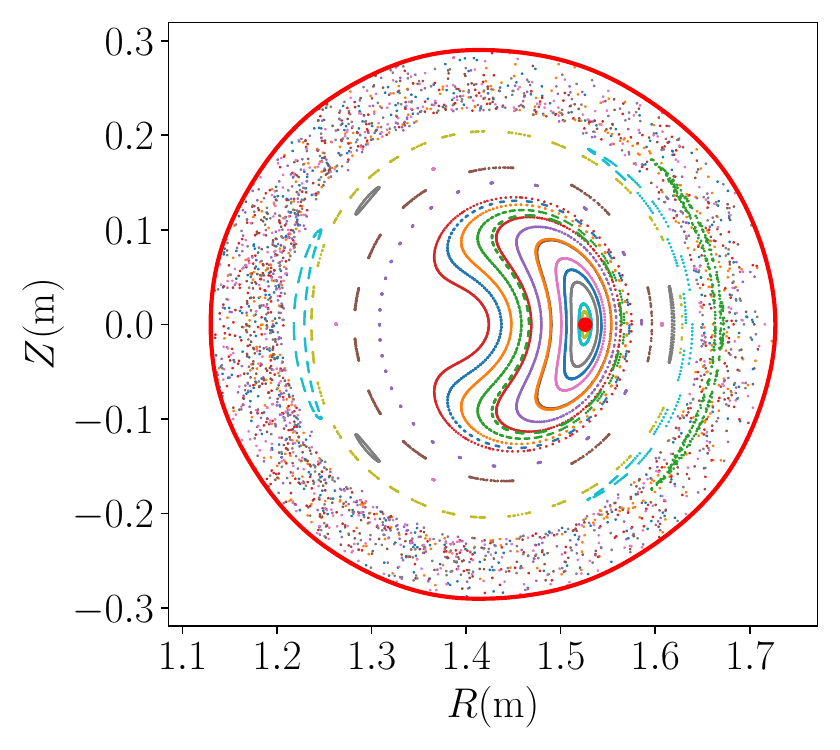}
        }\hfill
    \caption{Poincar\'{e} plot at $\zeta=0$ in the inner subvolume with different $\left<\mathcal{H}\right>$. The outmost red curve is the cross section of the interface. }
    \label{fig:poincare}
\end{figure}

Without the assumption of nested surfaces, there could be magnetic islands and chaotic fields. 
To enforce various boundary conditions, it is convenient to use toroidal coordinates $(s,\theta,\zeta)$, which are adapted to the plasma boundary and the interfaces. 
For stellarator symmetric equilibrium, the toroidal surface $\mathbf{x}_s$ can be described by $\mathbf{x}_s=R(\theta,\zeta)\mathbf{e}_R+Z(\theta,\zeta)\mathbf{e}_Z$ with
\begin{equation}
    R(\theta,\zeta) = \sum_m\sum_n R_{m,n}\cos(m\theta-n\zeta)  \ ,
\end{equation} 
\begin{equation}
    Z(\theta,\zeta) = \sum_m\sum_n Z_{m,n}\sin(m\theta-n\zeta)  \ ,
\end{equation}
where $\mathbf{e}_R$ and $\mathbf{e}_Z$ are the basis vectors in cylindrical coordinates. 
Similarly, the vector potential on a toroidal surface, $\mathbf{A}_s=A_{\theta}\nabla\theta+A_{\zeta}\nabla\zeta$,
\begin{equation}
    A_{\theta} = \sum_m\sum_n A_{\theta,m,n}\cos(m\theta-n\zeta)   \ ,
\end{equation}
\begin{equation}
    A_{\zeta} = \sum_m\sum_n A_{\zeta,m,n}\cos(m\theta-n\zeta)   \ .
\end{equation} 
As mentioned above, we use trigonometric function basis in the poloidal and toroidal directions to express physical quantities, and we use polynomial function basis in the radial direction to discretize these quantities. 

The input data in these simulations are toroidal flux $\Psi_i$, stepped pressure $p_i$ and helicity $\mathcal{H}_i$, where $i=1,2$ represent the inner and outer subvolume, respectively. 
The difference between the two stepped-pressures is set to $p_1-p_2=560\ \mathrm{Pa}$ and the total toroidal flux is $\Psi=\Psi_1+\Psi_2=0.1298\ \mathrm{Wb}$ according to the discharge and design parameters of KTX. 
The step of the pressure profile is determined by the magnetic flux of the subvolume. 
Based on previous results \cite{biewer2003electron, gobbin2011vanishing}, the transport barrier locates at a relatively outer position in the radial direction. 
In our simulaton, the flux of inner subvolume is $\Psi_1=0.86\Psi$, which keeps the position of the interface relatively out. 
The SPEC outputs are the shape of the interface and the vector potential $\mathbf{A}$. 
The magnetic field can be obtained by $\mathbf{B}=\nabla\times\mathbf{A}$.

In our simulations, we found different core magnetic helicity density $\left<\mathcal{H}\right>$ results in magnetic fields with different topologies. 
Here, core magnetic helicity density $\left<\mathcal{H}\right>$ is the average of the helicity in the inner subvolume,
\begin{equation}
    \left<\mathcal{H}\right> = \frac{\int_{\mathcal{V}_1}\mathbf{A}\cdot\mathbf{B}\mathrm{d}\mathcal{V}_1} {\int_{\mathcal{V}_1}\mathrm{d}\mathcal{V}_1} \ .
\end{equation}
The system is axisymmetric when the helicity is at a lower density. 
As $\left<\mathcal{H}\right>$ increases, the plasma spontaneously generates three-dimensional structures. 
The magnetic field gradually transforms from an axisymmetric state to DAx state and then to SHAx state.
\Fref{fig:flux surface} shows the flux surfaces of these different self-organized states. 
There is a new magnetic axis generated and the old magnetic axis disappears in the process of changing the magnetic topology. 
\Fref{fig:poincare} shows more details of the magnetic field through the Poincar\'{e} cross-section of six cases with different helicity densities. 
It should be noted that there is no Poincar\'{e} plot in the outer subvolume because the toroidal field is very small at the edge.

\section{Magnetic field properties of different self-organized states} \label{field}

We simulated a large number of self-organized states with $\left<\mathcal{H}\right>$ ranging from $5.2\times10^{-3}$ to $9.2\times10^{-3} \mathrm{T^2\cdot m}$ and analyzed the details of the magnetic field in different states. 
When $\left<\mathcal{H}\right>$ is less than $5.84\times10^{-3} \mathrm{T^2\cdot m}$, the system is always in the axisymmetrical state. 
If the core helicity density exceeds this threshold, there will be a new magnetic axis and the magnetic field will be in the DAx state where two flux tubes are twisted with each other. 
New flux tubes become larger as helicity increases. 
When $\left<\mathcal{H}\right> > 8.61\times10^{-3} \mathrm{T^2\cdot m}$, the old flux tube disappears, there is only a large flux magnetic tube and the plasma is in the SHAx state.
\Fref{fig:poincare} illustrates this regulation by six Poincar\'{e} plots with different core helicity density.

\begin{figure}
    \centering
    \includegraphics[width=0.7\textwidth]{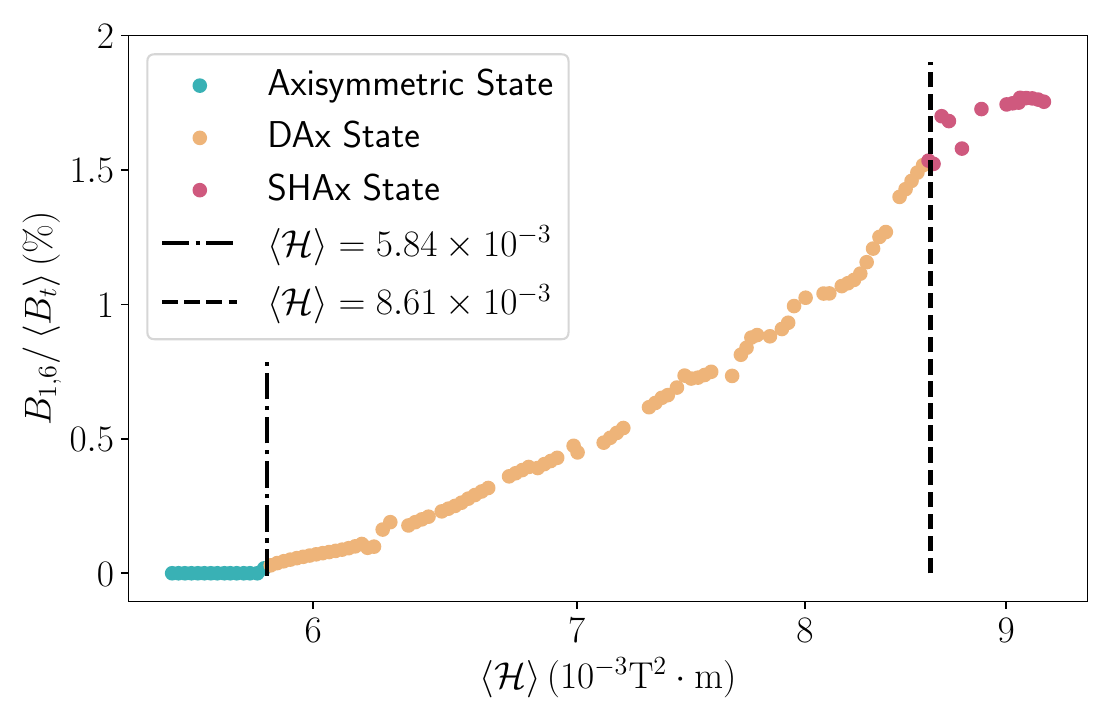}
    \caption{The relationship between the amplitude of the $m=1$, $n=6$ mode and the core helicity density $\left<\mathcal{H}\right>$. }
    \label{fig:dominantMode}
\end{figure}

A $m=1, n=6$ Fourier component of the magnetic field at the plasma edge becomes more and more evident as the core magnetic helicity density increases. 
\Fref{fig:dominantMode} shows the relationship between the amplitude of the $B_{1,6}$ normalized to the mean toroidal field and the core helicity density in different self-organized states. 
While the system is axisymmetric the amplitude of $B_{1,6}$ is always at a low level. 
When the magnetic field has a three-dimensional structure, whether DAx state or SHAx state, the $B_{1,6}$ component at the boundary is much greater than the value in the axisymmetric state. 
If $\left<\mathcal{H}\right>$ is large enough after the plasma spontaneously transforms into the SHAx state, the amplitude of the $B_{1,6}$ will reach the maximum.

\begin{figure}
    \centering
    \includegraphics[width=0.7\textwidth]{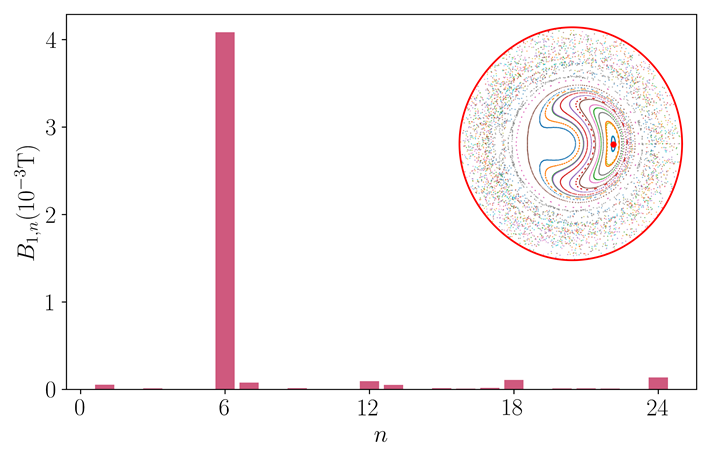}
    \caption{The Poincar\'{e} plot and $n$ spectrum of $m=1$ modes in a typical SHAx state. }
    \label{fig:mode}
\end{figure}
\begin{figure}
    \centering
    \includegraphics[width=0.7\textwidth]{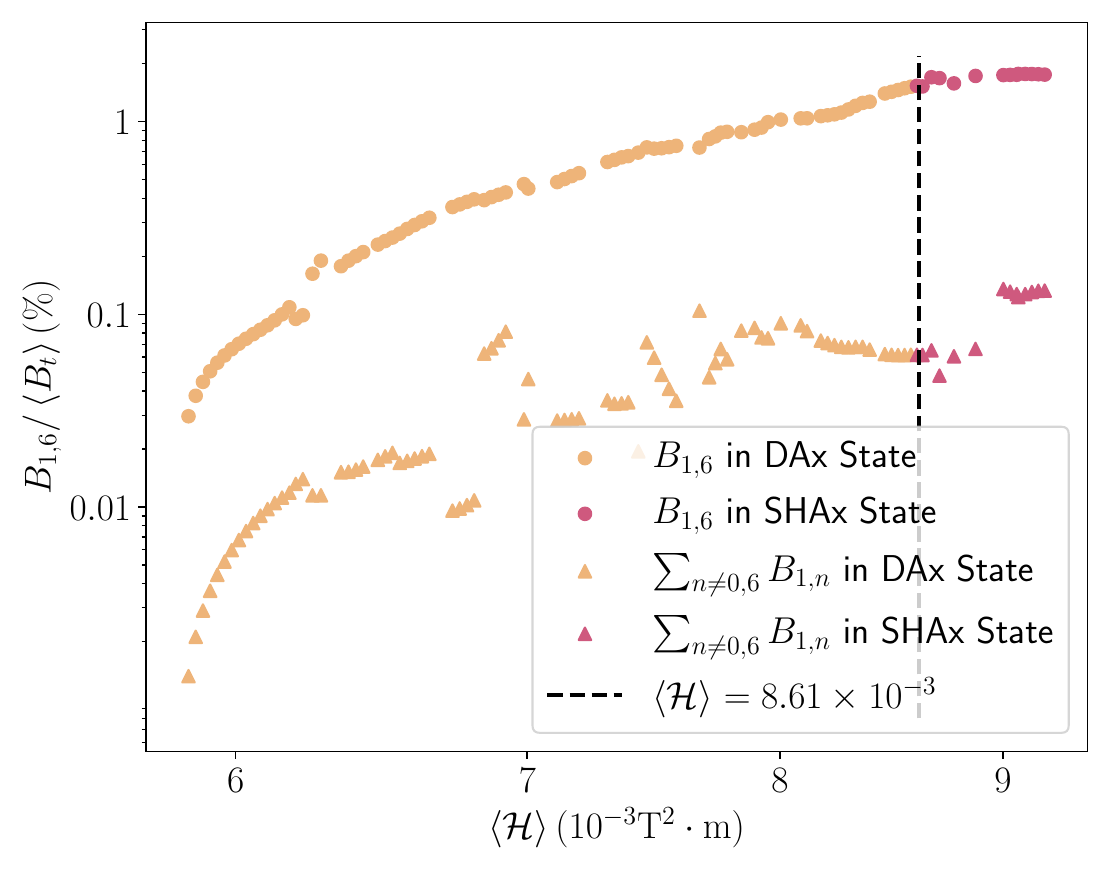}
    \caption{The amplitude of the dominant mode and the secondary mode as a function of $\left<\mathcal{H}\right>$. }
    \label{fig:QSHMode}
\end{figure}

When the plasma spontaneously generates a three-dimensional structure, $B_{1,6}$ as the dominant mode is more obvious than other secondary modes. 
\Fref{fig:mode} shows a Poincar\'{e} plot and $n$ spectrum of $m=1$ magnetic modes in a typical SHAx state and there is a $m=1, n=6$ dominant mode at the edge in the three-dimensional self-organized state of KTX. 
The same phenomenon has been experimentally discovered in other RFPs and there are different toroidal mode numbers of dominant mode because of the difference in minor and major radius. 
RFX-mod has a $m=1, n=7$ dominant mode and MST has a $m=1, n=5$ dominant mode \cite{lorenzini2009self, bergerson2011bifurcation, sarff2013overview}.
Although there is a dominant mode in our simulation, the other secondary modes will not disappear. 
\Fref{fig:QSHMode} shows that the amplitude of other $m=1$ secondary helical modes that still exist here is about 10$\%$ of the dominant mode.

\begin{figure}
    \centering
    \includegraphics[width=0.7\textwidth]{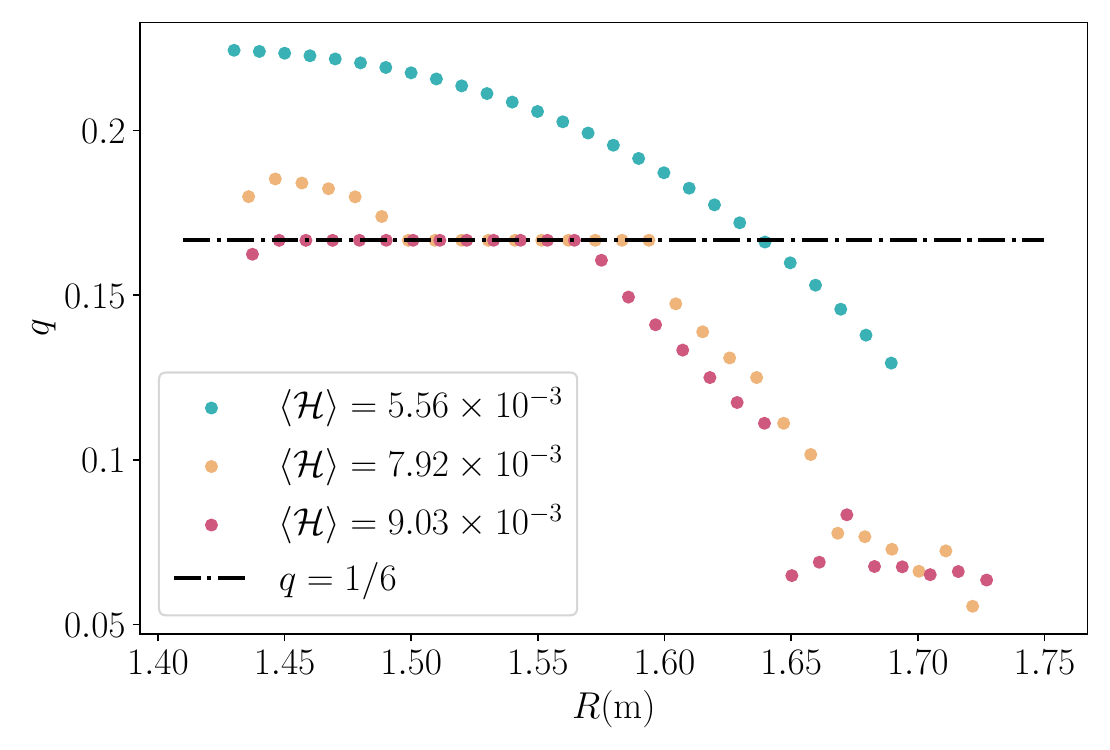}
    \caption{The safety factor of different self-organized states. }
    \label{fig:safetyFactor}
\end{figure}

The magnetic shear of the new flux tube in the DAx state and the core in the SHAx state will vanish. 
\Fref{fig:safetyFactor} shows the inner subvolume safety factor profiles of three self-organized states with different $\left<\mathcal{H}\right>$.
Both the DAx and SHAx states have lower safety factors than the axisymmetric state. 
The safety factor of the magnetic island in the DAx state and the core in the SHAx state is a constant of $1/6$. 
This shows that the helical magnetic field in the core of the SHAx state is a large magnetic island.

\begin{figure}
    \centering
    \includegraphics[width=0.7\textwidth]{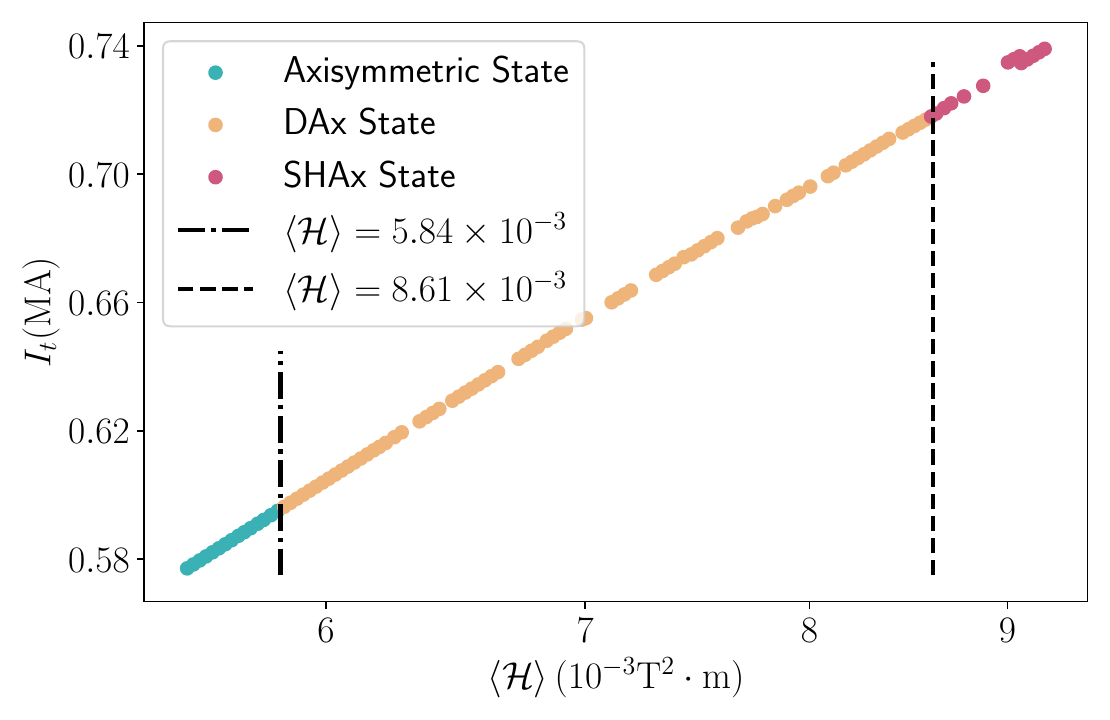}
    \caption{Plasma current as a function of $\left<\mathcal{H}\right>$. }
    \label{fig:current}
\end{figure}

There is an apparent positive linear correlation between $\left<\mathcal{H}\right>$ and the plasma current. 
In different self-organized states, the current is enhanced with increasing magnetic helicity. 
\Fref{fig:current} shows this relationship. 
The helicity flux density measurement system is implemented in KTX, and a similar linear relationship was obtained \cite{10.1063/5.0073486, Chen_2024}.
For KTX, when the average toroidal magnetic field $\left<B_t\right>\approx0.26\ \mathrm{T}$, we estimate that a plasma current of about  $0.6\ \mathrm{MA}$ is needed to spontaneously transform into a three-dimensional state and when the current reaches $0.72\ \mathrm{MA}$, the plasma can self-organize to form the SHAx state. 
There is also a higher probability of obtaining QSH spectra with higher plasma current experimentally \cite{martin2003overview, bolzonella2002magnetic, marrelli2002quasi, valisa2008high}.

\section{Decomposition of the magnetic helicity} \label{decomposition}

Mathematically, the sum of the Gauss linking number over every pair of field lines within a volume is the magnetic helicity \cite{moffatt_1969}. 
The linking number $\mathcal{L}_{\mathbf{xy}}$ of the two curves $\mathbf{x}$ and $\mathbf{y}$ is
\begin{equation}
    \mathcal{L}_{\mathbf{xy}} = 
    \frac{1}{4\pi}\oint_\mathbf{x}\oint_\mathbf{y}
    \frac{(\mathrm{d}\mathbf{x}\times\mathrm{d}\mathbf{y})\cdot(\mathbf{x}-\mathbf{y})}{|\mathbf{x}-\mathbf{y}|^3} \ . 
\end{equation}
If there are $N$ tubes and the flux carried by each tube is $\Psi_i$, $i=1, \dots, N$, we can calculate the sum of linking numbers with flux weighting. 
If we let $N\to\infty$ with $\Psi_i\to 0$, the tubes degenerate into field lines and we have
\begin{equation}
    \mathcal{H} 
    = \lim_{N\to\infty\atop \Psi\to 0} \sum_{i=1}^N\sum_{j=1}^N \mathcal{L}_{ij}\Psi_i\Psi_j
    = \int \mathbf{A}\cdot\mathbf{B} \mathrm{d}\mathcal{V} \ ,
\end{equation} 
where the Coulomb gauge is used in the vector potential.

\begin{figure}
    \centering
    \includegraphics[width=\textwidth]{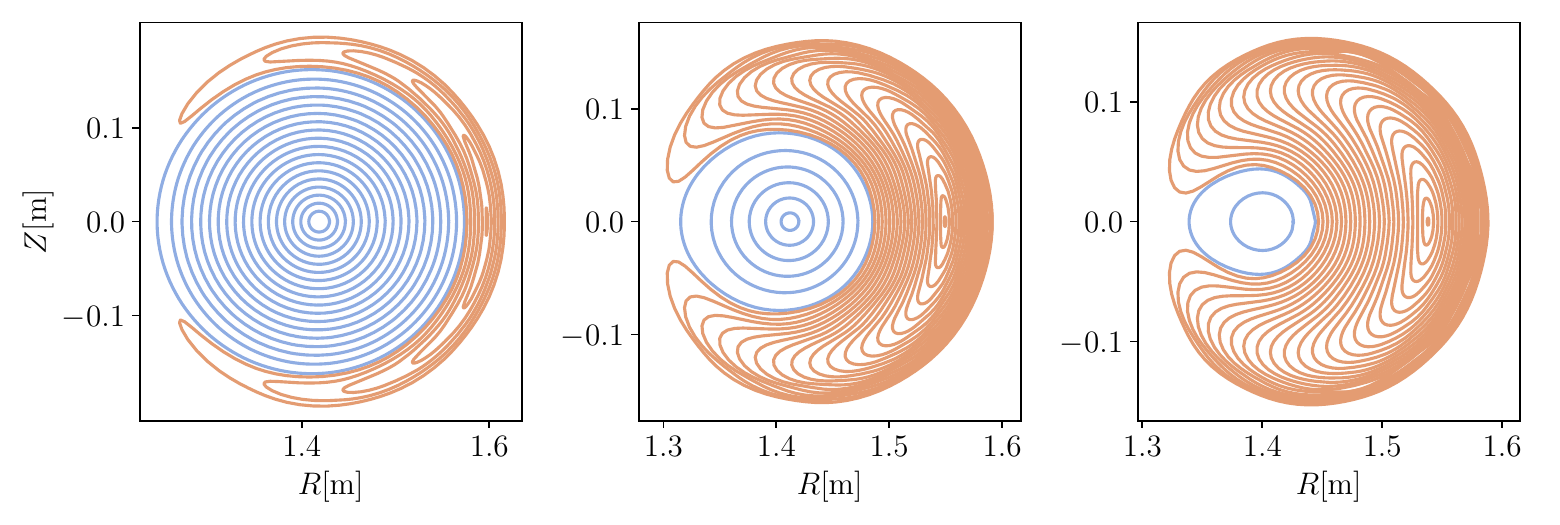}
    \caption{The growth of the new flux tube (orange) and the reduction of the old flux tube (light blue). }
    \label{fig:twoTubes}
\end{figure}

Helicity measures the link and twist between magnetic field lines together with the writhing of the field lines themselves \cite{berger_field_1984, Mitchell_A_Berger_1999, CompleteMeasurement}. 
We can analyze the components of the helicity at different states in RFP.
As shown in \Fref{fig:twoTubes}, the transition of these self-organized states is accompanied by the generation and growth of the new flux tube $\Psi_\mathrm{n}$, and the reduction and disappearance of the old flux tube $\Psi_\mathrm{o}$. 
We can use the mutual helicity to describe the link between these two tubes 
\begin{equation}
    \mathcal{H}_\mathrm{mutual} = 2\mathcal{L}\Psi_\mathrm{o}\Psi_\mathrm{n} \ . 
\end{equation} 
The self helicity within the flux tube can be calculated  
\begin{equation}
    \mathcal{H}_\mathrm{self} = 2\int_0^\Psi\iota\Psi\mathrm{d}\Psi \ ,
\end{equation}
where $\iota$ is the rotational transform and the average Gauss linking number of the field line and the axis in one toroidal period.
According to the C\v{a}lug\v{a}reanu invariant \cite{calugareanu1959integrale, white1968self}, the linking and self helicity can be produced by the writhing of the axis itself and the twisting of the field lines around the axis. 
The writhing helicity $\mathcal{H}_{\mathrm{writhing}}=\mathrm{Wr}\Psi^2$ measures the three-dimensional structure of the magnetic axis with the writhing number 
\begin{equation}
   \mathrm{Wr} = \frac{1}{4\pi}\oint_\mathrm{Axis}\oint_\mathrm{Axis}
    \frac{(\mathrm{d}\mathbf{x}\times\mathrm{d}\mathbf{y})\cdot(\mathbf{x}-\mathbf{y})}{|\mathbf{x}-\mathbf{y}|^3} \ .
\end{equation}
The twisting helicity $\mathcal{H}_{\mathrm{twisting}}=\mathcal{H}_{\mathrm{self}}-\mathcal{H}_{\mathrm{writhing}}$ reflects the contribution from the twisting of field lines around the axis.

\begin{figure}[htbp]
    \centering
    \subfloat[mutual helicity and self helicity]{
        \label{fig:mutual-self}
        \includegraphics[width=0.7\linewidth]{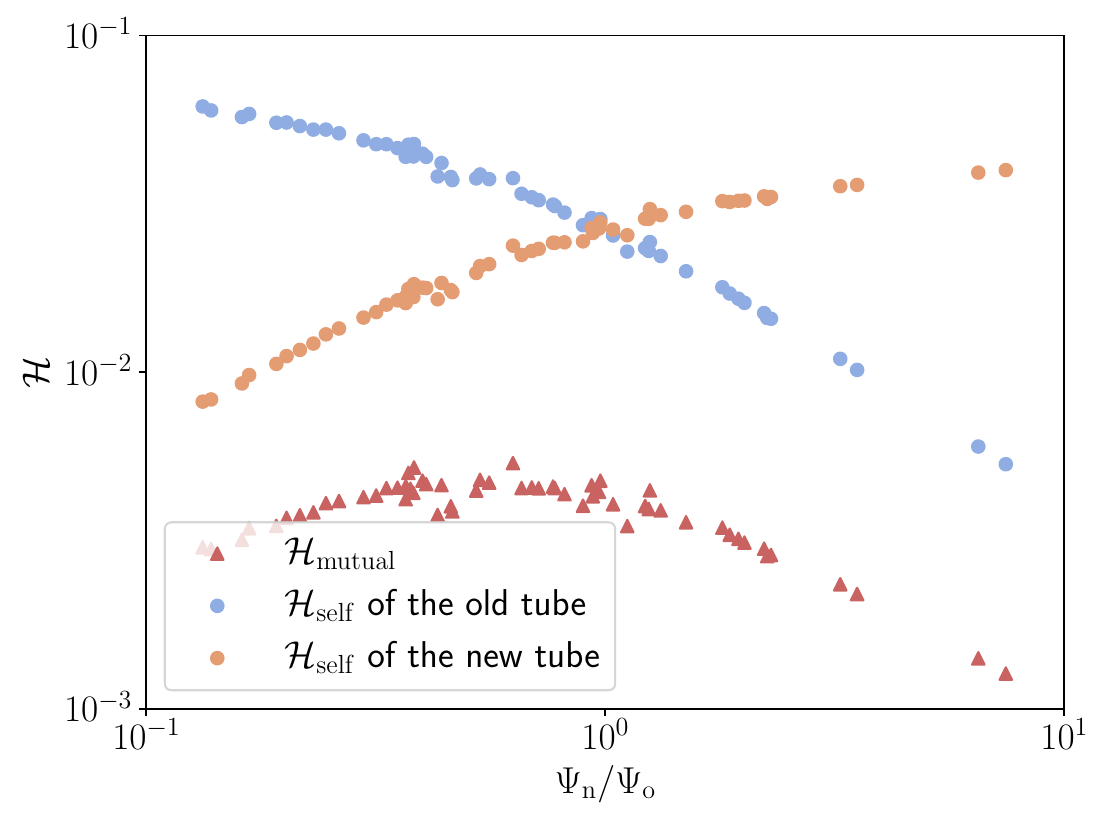}
        }\hfill
    \subfloat[twisting helicity and writhing helicity]{
        \label{fig:twisting-writhing}
        \includegraphics[width=0.7\linewidth]{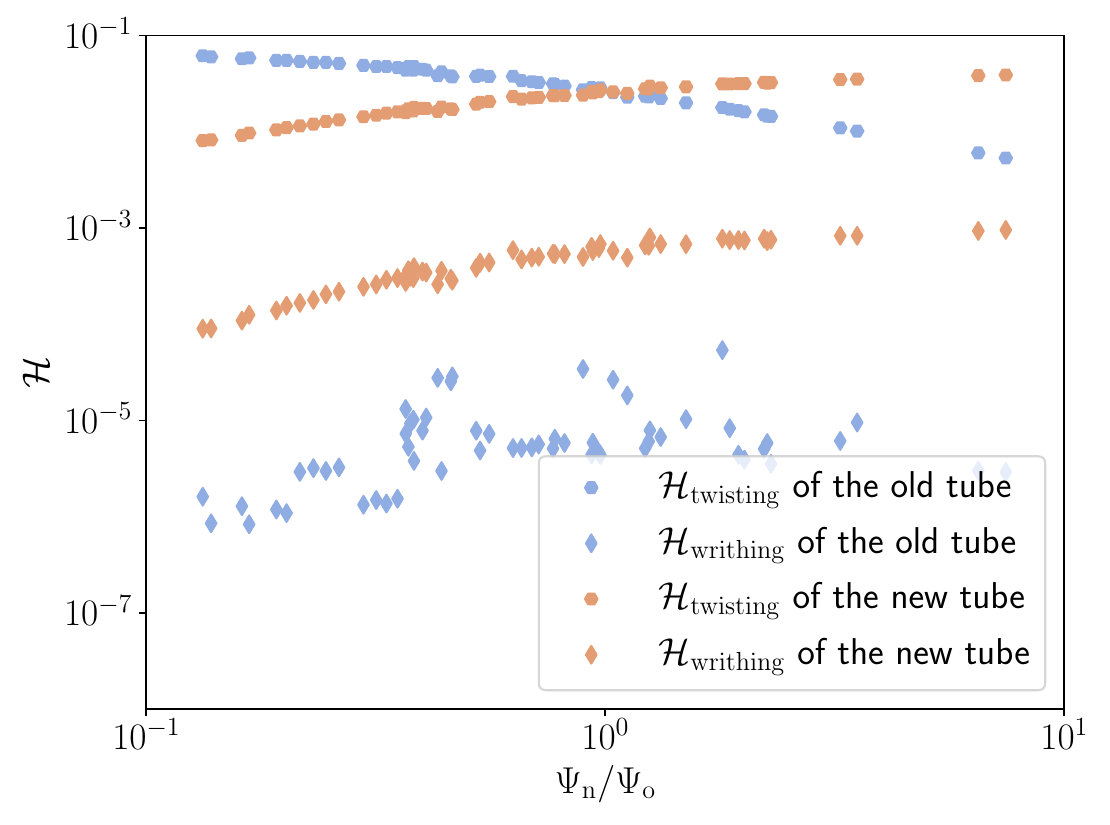}
        }
    \caption{Helicity decomposition in different DAx states. $\Psi_{\mathrm{n}}/\Psi_{\mathrm{o}}$ is the magnetic flux ratio of the new flux tube to the old tube }
    \label{fig:decomposition}
\end{figure}

\Fref{fig:decomposition} illustrates the decomposition of helicity in different DAx states.
Note that when calculating magnetic helicity, we ignore chaotic regions. 
With the changes in the flux surfaces, helicity components are also varying.
As shown in \Fref{fig:mutual-self}, the $\mathcal{H}_\mathrm{self}$ of the old flux tube gradually decreases and the $\mathcal{H}_\mathrm{self}$ of the new flux tube gradually increases, both of which are significantly larger than the mutual helicity $\mathcal{H}_\mathrm{mutual}$. 
\Fref{fig:twisting-writhing} describes more details of the twisting and writhing components in self helicity. 
The magnetic axis of the new flux tube has a significant three-dimensional structure, thus its writhing helicity $\mathcal{H}_\mathrm{writhing}$ is much larger than that of the old flux tube. 
The $\mathcal{H}_\mathrm{self}$ of the old flux tube mainly consists of the twisting helicity $\mathcal{H}_\mathrm{twisting}$, while the writhing helicity $\mathcal{H}_\mathrm{writhing}$ is small enough to be ignored.

\section{Summary and discussion} \label{summay}

Different self-organized states on KTX were simulated using the SPEC code. 
We use the barrier as an interface to divide the plasma into two subvolumes, and the toroidal flux in these two subvolumes remains unchanged, ensuring that the barrier is always in a relatively out position. 
The relaxed state in each region minimizes the energy functional under the local helicity constraint.
We reproduced different self-organized states by changing the core helicity density, which plays an important role in the plasma relaxed state.
The system is axisymmetric with low helicity and as the core helicity density $\left<\mathcal{H}\right>$ increases, the self-organized state of the plasma gradually transforms into the DAx state and then into the SHAx state.

Beyond its three-dimensional geometric structure, the high $\left<\mathcal{H}\right>$ self-organized state exhibits distinct magnetic field properties. 
The high helicity makes the dominant mode at the plasma edge more pronounced, characterized by a toroidal number of 6.
The emergence of a three-dimensional structure eliminates magnetic shear in the core, corroborating experimental observations on other RFP devices.  
Additionally, we found an approximate linear relationship between the plasma toroidal current and the core helicity density.
With this information, we predict that the KTX plasma will enter the DAx state and have three-dimensional structures when the plasma current is higher than $0.6\ \mathrm{MA}$.
Furthermore, when the plasma current reaches $0.72\ \mathrm{MA}$, the plasma will spontaneously transform into the SHAx state.

We further analyzed the components of magnetic helicity.
Changes in magnetic topology are accompanied by the generation of a new flux tube and the disappearance of the old flux tube. 
The total helicity is primarily composed of the self-helicity within these two flux tubes, rather than the mutual helicity.
The helicity of the old flux tube was predominantly characterized by twisting, while the new flux tube has a certain writhing helicity with an evident three-dimensional structure. 

The MRxMHD model and the stellarator equilibrium code SPEC reveal essential equilibrium properties of RFP.
SPEC demonstrates its capability for RFP devices and can be used in other RFP applications, such as equilibrium reconstructions.
In the future, numerical results can be further validated through experiments.

\section*{Acknowledgments}\label{acknowledgments}

The authors would like to acknowledge Dr. Stuart Hudson and the SPEC developers team for supporting the use of SPEC.
KL and CZ would like to thank Dr. Haifeng Liu and Dr. Peifeng Fan for fruitful discussions.
This work was supported by the University of Science and Technology of China and the National Natural Science Foundation of China.
WM and GZ were supported by the National Natural Science Foundation of China under grant No. 12175227.
Numerical calculations were performed at Hefei Advanced Computing Center.

\section*{References}
\bibliographystyle{unsrt}
\bibliography{ref}

\end{document}